# Tunneling of Elastic Waves in a Tapered Waveguide

Alexandre Yoshitaka Charau,[1,2,*] Jérôme Laurent,[1,†] and Tony Valier-Brasier[2,‡]

[1]*Université Paris-Saclay, CEA, List,* F-91120, Palaiseau, France
[2]*Institut Jean Le Rond ∂'Alembert, Sorbonne Université,*
*UMR CNRS7190, 4 Place Jussieu, Paris 75005, France*

Understanding how evanescent modes mediate energy transfer in tapered elastic waveguides is of paramount interest, as it unlocks new strategies for wave control and manipulation. Evanescent modes play a crucial role in energy localization and in the emergence of thickness resonances. We report the first unambiguous experimental evidence of Lamb mode tunneling near turning points, revealing how energy can traverse an evanescent barrier and recover its propagative nature after a finite transit time. Focusing on waveguides with linearly varying thickness, we show that the $S_2$-mode becomes evanescent over a narrow frequency band, enabling a tunneling-like phenomenon. Our study demonstrates that the barrier width is governed by the elastic properties of the material, particularly the Poisson's ratio, within a confined range bordering the Dirac cone. Numerical results exhibit excellent agreement with predictions from the Wentzel-Kramers-Brillouin approximation. These findings provide compelling evidence that, for a specific barrier width, evanescent modes mediate energy transfer across regions classically forbidden to propagating waves, revealing the mechanisms governing transmission, localization, and mode conversion in structured or corrugated elastic waveguides.

*Introduction*—The physical phenomenon of crossing a classically forbidden region was first observed in optics. In 1730, Newton described a pilot experiment demonstrating Frustrated Total Internal Reflection (FTIR), in which a light beam tunnels between two prisms separated by a narrow gap [1–4]. This intriguing effect—the penetration of the evanescent wave into the second medium during total internal reflection—was later studied by Fresnel, Young, Huygens, Rayleigh, Bose, and others. However, the theoretical notion of tunneling and the term itself first appeared in quantum mechanics, where it refers to the phenomenon in which a particle crosses a potential barrier despite lacking sufficient energy according to classical theory. A comparable mechanism arises in classical wave physics, where evanescent fields enable energy transmission through regions that do not support conventional wave propagation. Such behavior has been then observed in various physical systems, including electrical circuits [5], electromagnetic [6–9], phononic crystals [10, 11], metamaterials [12], and acoustics waves [13–16]. Within the context of elastic waveguides exhibiting geometric or material inhomogeneities, distinctive phenomena emerge from the interaction between propagating and evanescent Lamb modes [17]. While propagating modes are responsible for long-range energy transport, evanescent modes, although spatially confined, play a crucial role in mediating localized interactions. These interactions become particularly significant near discontinuities or regions of rapid structural variation, where they can give rise to abnormal wave phenomena such as mode conversion, negative refraction, and negative reflection [18–22] or as Maxwell's fish-eye lens [23]. Because energy becomes trapped-like in these evanescent zones, such interactions also contribute to the emergence of thickness resonances.

In this context, few intriguing experimental studies have reported observations reminiscent of tunneling-like behavior. Notably, Alippi *et al.* [24] investigated the transmission of Lamb modes through a locally thinned region and interpreted the apparent propagation of the $S_1$-mode across this classically forbidden section as an "acoustic tunneling" process, with traversal times seemingly independent of the barrier length, in analogy with the Hartman effect [25, 26]. This interpretation, later upheld by Germano *et al.* [27], was explicitly inspired by tunneling phenomena in quantum mechanics and electromagnetism [28, 29]. However, subsequent investigations have shown that this behavior can be entirely understood within the framework of classical elastodynamics. Finite-element (FE) simulations by Kuznetsova *et al.* [30] demonstrated that the apparent "instantaneous transmission" across a groove is not a manifestation of tunneling, but results from mode conversion at the groove edges. The incident $S_1$-mode is converted into lower-order modes capable of propagating across the thinner section, and subsequently reconverted into the $S_1$-mode beyond the groove. This mechanism fully accounts for the observed transmission within the framework of classical elastodynamics, without invoking any noncausal or quantum-analog processes. The pulse traversal time is therefore finite and merely determined by the propagation velocities of the converted modes, in complete agreement with conventional elastic wave theory.

In elastic plates with gradually varying thickness, Lamb modes evolve adiabatically [31–33], continuously adjusting their wavenumbers to follow the local dispersion characteristics [34]. As the plate thickness approaches the thickness for a given mode, total reflection is typically expected to occur, preventing further propagation. Yet, experimental observations have revealed par-tial conversions into lower-order modes near these transitional regions, implying that mode coupling in this regime may involve mechanisms beyond standard adiabatic theory introduced by [35]. Unlike phenomena relying on band gaps in phononic crystals [36], the anomalous transmission observed here does not stem from periodic structuring but instead arises from the presence of a slowly varying cross section along the length of the plate.

In this Letter, we experimentally report that the propagating second-order symmetric Lamb mode ($S_2$) can traverse a classically forbidden region in an elastic waveguide with linearly varying thickness. This behavior, manifesting as transmission through an evanescent barrier, highlights a tunneling-like phenomenon analogous to quantum mechanical tunneling, yet firmly rooted in classical elastodynamics. To elucidate the parameters governing this counterintuitive transmission, we perform a comprehensive parametric study using numerical simulations based on the scattering-matrix method [37]. Our analysis reveals that the geometry of the thickness gradient and the intrinsic elastic properties of the material, particularly the Poisson's ratio $\nu$, concomitantly govern the transmission and reflection behavior of the mode.

*Complex spectra of Lamb's modes*—Figure (1) presents the complex solutions $k(f)$ of the Rayleigh-Lamb equation for a steel plate with thickness $h = 1$ mm, where the wavenumber is expressed as $k = k' + ik''$, with $k'$ and $k''$ denoting the real and imaginary parts, respectively. The solutions are shown for the $S_1$- and $S_2$-modes [38].

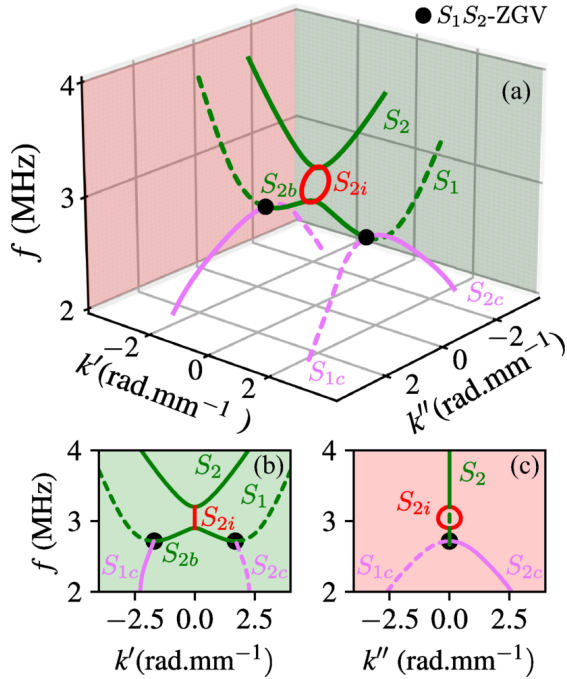

FIG. 1. Theoretical $S_1$- and $S_2$-mode dispersion curves for a steel plate ($\nu = 0.29$) with 1 mm thickness: (a) complex wavenumber plane; (b) real plane; (c) imaginary plane. Green curves are for propagative mode, red for evanescent modes and purple for inhomogeneous modes

At high frequencies, the $S_2$-mode initially exhibits a purely real solution, indicating propagative behavior. As the frequency decreases toward its value, $f_{c_{S_2}} = V_T/h$, where $V_T$ is the transverse wave velocity, the mode becomes evanescent, acquiring a purely imaginary wavenumber ($S_{2i}$). It remains in this evanescent state until the frequency reaches $f_{c_{S_{2b}}} = V_L/2h$, with $V_L$ the longitudinal wave velocity, at which point it reenters a propagative regime with a negative phase velocity ($S_{2b}$ backward mode). As the frequency decreases further, the mode converges toward the $S_1$-branch at the point where the $S_{2c}$ and $S_{1c}$ branches intersect, corresponding to the presence of an $S_1 S_2$ Zero-Group Velocity [39] (ZGV) mode [40].

*Experimental setup and results*—While this type of transition is usually described in terms of decreasing frequency, it can also be observed spatially by tapering the waveguide thickness at a fixed excitation frequency ($f_0$), as depicted in Fig. (2a). To investigate this effect, we use an aluminum plate with a linearly varying thickness from $h_1 = 4.5$ mm to $h_2 = 2.5$ mm over a length of $L = 250$ mm, corresponding to a dihedral angle $\theta = \arctan[(h_1 - h_2)/L] \approx 0.46°$. A longitudinal transducer is mounted on a tunable Perspex wedge to excite specific Lamb modes from the thick end of the plate. A wedge angle of $\alpha = 10°$ and a 20-cycle sine tone burst centered at 0.92 MHz apodized by a Hanning window are used to preferentially generate the propagating $S_2$-mode. The longitudinal and transverse wave velocities of the aluminum plate are measured as $V_L = 6.32$ mm/$\mu$s and $V_T = 3.12$ mm/$\mu$s ($\nu = 0.32$), respectively. These values are obtained using a laser ultrasonic technique based on the identification of ZGV-modes [41]. A scan is made using a multichannel random quadrature laser interferometer (Quartet, Sound&Bright) to measure the normal surface displacement along a 140 mm portion of the plate, with a spatial resolution of 0.4 mm. At the same time, the local thickness profile is recorded using a laser telemeter. Figures (2b) and (2c) show the theoretical spatial evolution of the real and imaginary components of the local wavenumber associated with the $S_2$-mode. As the wave propagates through the tapered region, the local wavenumber of the $S_2$-mode transitions from real to purely imaginary and then back to real in the vicinity of turning points.

To visualize this transition, the local wavenumbers are examined as a function of time ($t$). To isolate the frequency-dependent spatial characteristics of the wavefield, we perform a two-dimensional short-time Fourier transform (STFT). First, a temporal STFT is applied to the measured signal matrix $s(x,t)$ using a sliding time window, allowing extraction of the component at a specific temporal frequency $f_0$. Next, a spatial STFT is performed on this filtered signal, yielding the matrix $S_{f_0}(x,t,k)$, which describes the local wavenumber content at $f_0$ as a function of position and time. Figure (3) shows the temporal evolution of the local wavenumbers at $f_0 = 0.92$ MHz. The real part of the spectrograms is displayed to emphasize the spatial phase of the wave-field. In Fig. (3a), the $S_2$-mode is successfully excited and propagates forward through the thicker region of the plate. In Fig. (3b), as the mode enters the evanescent region, the spatial phase becomes constant, indicating that the real part of the local wavenumber drops to zero. At this stage, the wave begins to split: part of the energy is reflected, while part continues into the evanescent zone. In Fig. (3c), the splitting becomes evident: the reflected part propagates backward as the $S_2$-mode with a negative wavenumber, while the transmitted part continues forward as the $S_{2b}$-mode. Although the backward mode exhibits a negative wavenumber, it still propagates in the positive $x$ direction due to its positive group velocity. These observations confirm the presence of reflection associated with the thickness of the $S_2$-mode. However, in addition to the reflected component, a portion of the wave is transmitted beyond this region.

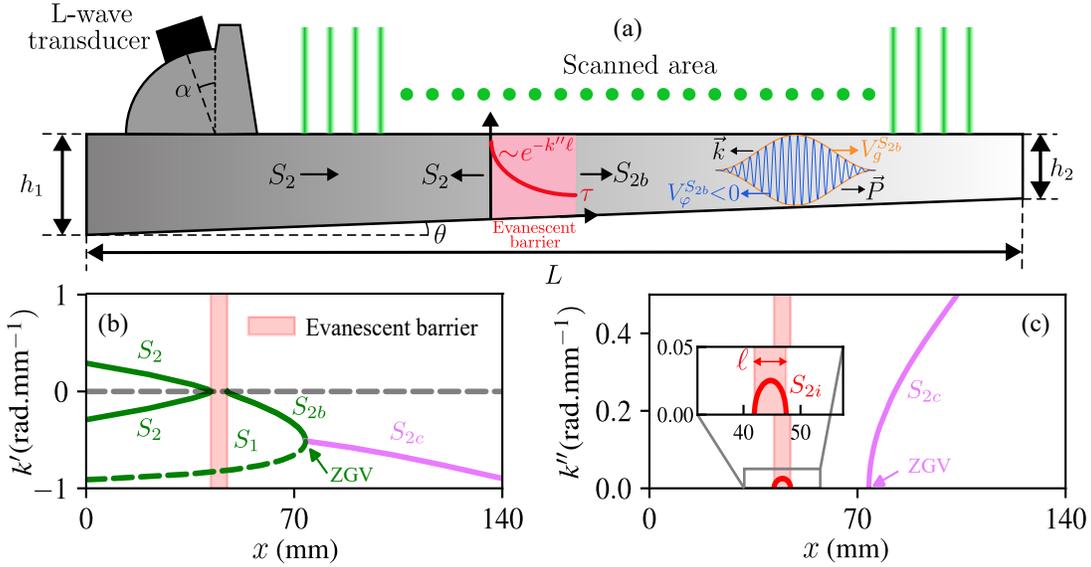

FIG. 2. (a) Schematic of the experimental setup. The evanescent barrier, highlighted by the red zone, is characterized by the length $\ell$ and the transit time $\tau$. A wave packet exhibiting a negative phase velocity ($V_\varphi^{S_{2b}}$), with the Poynting vector ($\vec{P}$) and wave vector ($\vec{k}$) oriented in opposite directions, is associated with the $S_{2b}$-mode that emerges immediately after the barrier. (b) Theoretical real local wavenumber $k'(x)$ and (c) imaginary local wavenumber $k''(x)$ over the scanned area.

The evanescent zone thus acts as a barrier $\ell$ that only partially blocks the wave, allowing a finite amount of energy to tunnel through. This behavior is analogous to a tunneling effect, where transmission occurs through a classically forbidden region due to the evanescent nature of the mode. The spatial extent of this barrier can be quantitatively estimated from the cutoff conditions of the involved Lamb modes, which depend on the longitudinal and transverse bulk velocities $V_L$ and $V_T$, respectively. It can be expressed as $\ell = x\bigl(h = V_L/(2f)\bigr) - x\bigl(h = V_T/f\bigr)$, where $x$ denotes the position along the length of the plate. In our configuration, this relation yields a barrier length of approximately 5 mm. We may further remark that, beyond the evanescent barrier, the $S_{2b}$-mode displays a negative phase velocity. Such a reversal of phase propagation is highly atypical in acoustic tunneling field and highlights the unusual wave dynamics that emerge near

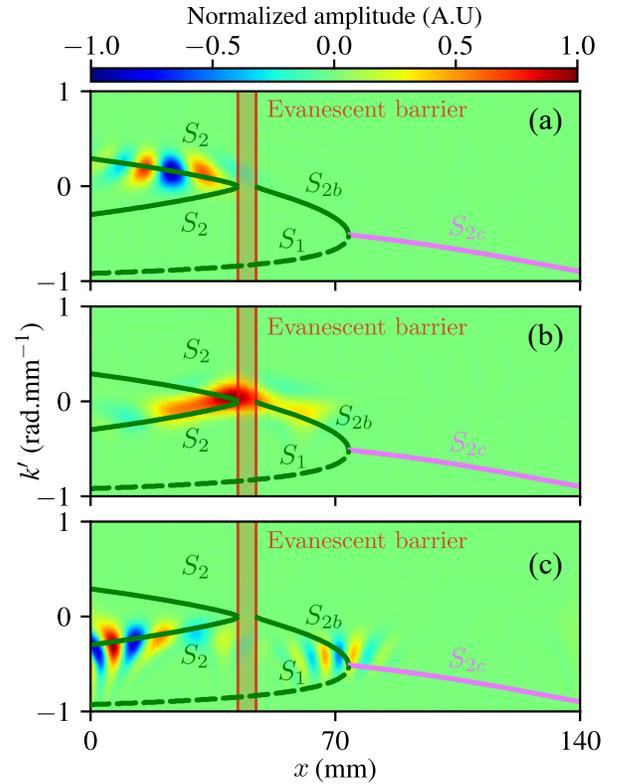

FIG. 3. Experimental spatial spectrogram $S_{f_0}(x,t,k)$ of the $S_2$-mode at different stages: (a) before tunneling-like at 58 $\mu$s; (b) interaction with the evanescent region at 71 $\mu$s; (c) after tunneling-like at 83 $\mu$s. The real parts of the local wavenumbers, derived from the theoretical dispersion curves, are overlaid as solid lines to highlight the transition between propagating and evanescent regimes.

the barrier termination. After crossing the ZGV point,

the $S_1$-mode also reverses its direction of propagation like $k' < 0$, completing this remarkable sequence of backward wave phenomena. Also, to further illustrate the dependence of the tunneling mechanism on the tapering direction, we analyze in the Supplemental Material [42] a "reversed-gradient" configuration within the barrier in which the thickness increases across the evanescent region. In this configuration, which is similar to the case examined by Nielsen and Peake [14], the reverse thickness gradient leads to the reemergence of the $S_2$-mode with a positive wavenumber beyond the barrier.

*Numerical and theoretical predictions*—To investigate how material properties influence the tunneling-like behavior of the $S_2$-mode, we perform a numerical parametric study focused on the effect of Poisson's ratio ($\nu$). The evanescent barrier that forms at a fixed frequency is governed by the cutoff thicknesses of the $S_2$- and $S_{2b}$-modes, which are directly related to the elastic wave velocities $V_L$ and $V_T$. These velocities define $\nu$ via the relation $\nu = (1 - 2\kappa^2)/[2(1 - \kappa^2)]$, where $\kappa = V_T/V_L$ [48]. To isolate the influence of $\nu$, we arbitrarily fix the transverse velocity $V_T$ at 3.10 mm/$\mu$s and vary the longitudinal velocity $V_L$ from 5.6 to 6.8 mm/$\mu$s in 25 discrete steps. This variation yields a corresponding range of Poisson's ratio values from 0.279 to 0.368. Surprisingly, this effect is confined to a narrow slice of less than 18% of the ordinarily material range of Poisson's ratio from $0 \leqslant \nu < 0.5$ (either, $0 \leqslant \kappa < 1/\sqrt{2}$). The simulations are conducted using the `CIVA` software, which implements a Hybrid Semi-Analytical Finite Element/Finite Element (SAFE/FE) model. This model enables the analysis of wave propagation in structures with nonuniform geometry and allows for the calculation of the scattering matrix, from which we extract the energy partition between the incident, reflected, and transmitted modes in a linearly tapered waveguide matching the configuration shown in Figure (2).

Consistent with the semianalytical solutions of the transcendental equation, in the Supplemental Material [42], we demonstrates that the evanescent barrier length ($\ell$) can be also predicted heuristically using an asymmetric semi-elliptical function depending on $\nu$, $f$ and $\theta$. Figures 4(a) and 4(b) illustrate how $\nu$ influences the imaginary part of the local wavenumber and the barrier width $\ell$. As $\nu$ increases up to $1/3$, both the imaginary wavenumber and the barrier width $\ell$ decrease and approach zero. Beyond this point, both quantities begin to increase again. This behavior can be explained by the fact that the cutoff thicknesses of the $S_2$ and $S_{2b}$ modes coincide, and the dispersion relation exhibits a linear crossing characteristic of a Dirac cone [49–51]. As a result, no evanescent region is formed, and the mode transitions directly between propagating states without the need for tunneling. This is also reflected in Figure (4c), where the numerically computed reflected and transmitted energies of the $S_2$-mode are shown as scatter plots. At $\nu = 1/3$, the transmission reaches its maximum while the reflection drops to zero, clearly confirming the disappearance of the evanescent barrier. In contrast, for lower Poisson's ratios, such as $\nu = 0.3$, the $S_2$ and $S_{2b}$ cutoffs are further apart, leading to a broader evanescent region and a higher imaginary wavenumber. These conditions amplify reflection and suppress transmission, highlighting the role of both $\ell$ and $k''$ in shaping the tunneling response. Crucially, since these quantities can be inferred from the theoretical dispersion curves and the known thickness profile of the waveguide, the tunneled energy can be predicted analytically. In the limit of a slowly varying thickness, the transmitted amplitude $C$ can be estimated using the Wentzel–Kramers–Brillouin (WKB) asymptotic approximation as $C = A\,e^{-\int_0^\ell k''(x)\,dx}$, where $A$ is the $S_2$-mode incident amplitude. By expressing the transmitted amplitude $C$ relative to the incident amplitude, the transmitted energy through the evanescent barrier can be approximated as $\mathbf{T} = (C/A)^2$. Consequently, the reflected energy is given by $\mathbf{R} = 1 - \mathbf{T}$. This analytical prediction was evaluated across the full range of Poisson's ratio values by integrating the local imaginary wavenumber $k''(x)$ obtained from the dispersion curves.

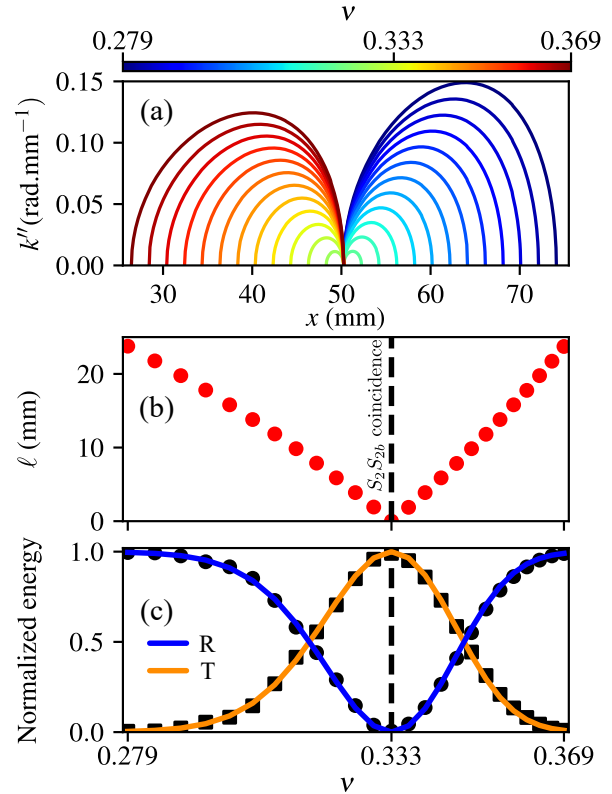

FIG. 4. Parametric analysis of the $S_2$-mode across an evanescent barrier as a function of Poisson's ratio ($\nu$) at 0.89 MHz: (a) Imaginary part of the local wavenumber $k''$; (b) Length $\ell$ of the evanescent barrier; (c) Normalized energy of the $S_2$-mode: transmitted (**T**) and reflected (**R**) energies as functions of $\nu$. Numerical results are shown as black squares (■) for **T** and black circles (●) for **R**, while WKB analytical predictions are plotted as solid orange and blue lines for **T** and **R**, respectively. Black dashed line denotes $\nu = 1/3$, corresponding to the occurrence of the Dirac cone.

When compared with the numerically computed en-ergies shown in Figure (4c), the results show excellent agreement. This confirms that the transmitted portion of the $S_2$-mode indeed undergoes an exponential decay characteristic of evanescent behavior, and that the energy partition between transmission and reflection is governed by the geometry and material parameters of the barrier. The strong consistency between theory and simulation supports interpreting this phenomenon as a classical analog to quantum mechanical tunneling. Just as a quantum particle can traverse a potential barrier despite having insufficient energy, the $S_2$-mode can penetrate the evanescent region, with its transmission governed by the effective height and width of the barrier, quantified here through the spatial distribution of $k''(x)$. In Supplemental Material [42], we present the time evolution of the wavenumber $k'(t)$, showing that the simulated dynamics closely match the experimental observations. We also highlight the $S_2$-mode transition between regimes of total reflection, total transmission near the Dirac cone, and partial transmission (tunneling), which can be tuned by adjusting $\nu$, taper slope, or excitation frequency ($f_0$). Finally, we report the apparent transit time ($\tau$) of the $S_{2i}$-mode energy confined within the evanescent barrier, which is neither barrier-length independent nor subject to Hartman-type saturation, as shown in the End Matter.

*Conclusion*—We have reported the first experimental evidence of Lamb mode tunneling near the turning points. We have experimentally and numerically demonstrated that the $S_2$-Lamb-mode can undergo a tunneling-like transition across an evanescent region in a smoothly tapered elastic waveguide. The transmission and reflection behaviors are governed by the local imaginary wavenumber or the width of the evanescent barrier, which depend sensitively on a narrow range of Poisson's ratio near the Dirac cone. The WKB approximation accurately predicts the transmitted and reflected energies, offering a robust analytical framework for this classical analog of quantum tunneling. These results, obtained under the assumption of a perfect waveguide (without considering viscoelastic attenuation), reveal a mechanism that could be exploited to control guided elastic waves. In particular, the combined effect of the evanescent barrier and the tapered geometry of the waveguide may give rise to acoustic diode-like (nonreciprocal) behavior [52], selectively transmitting or blocking the $S_2$-mode depending on the waveguide's thickness gradient, material properties, or excitation frequency. Lastly, extending such approach to Functionally Graded Materials (FGM), soft waveguides [53], and time-varying media [54] would be a promising direction for future work. Analogous tunneling effects may also arise in other mode transitions or across higher-order branches, involving different mode-switching pathways. Configurations capable of generating a barrier with a nonzero real part of the wavenumber, thus supporting an inhomogeneous mode with simultaneous exponential decay and spatial oscillation, represent a particularly promising avenue for future research. Equally compelling is the negative–refraction phenomenon that may arise once the barrier has been crossed [22, 55], effectively turning the variable-thickness region into a lens-like structure.

*Acknowledgments*—The authors would like to thank K. Jezzine and V. Baronian for insightful discussions and for implementing various upgrades to the "Guided Wave" module, available in the `CIVA software`.

*Data availability*—The data that support the findings of this article are not publicly available. The data are available from the authors upon reasonable request.

# End Matter

*Transit time within the evanescent barrier*—The apparent transit time $\tau$, which captures how long the wave energy is effectively "stored" within the evanescent barrier, is extracted from the peak arrival times of the wave–packet envelope on either side of the barrier, with the temporal resolution set by the sliding STFT window (6 $\mu$s wide). This procedure provides a reliable estimate of the traversal delay associated with the $S_{2i}$ evanescent mode.

Dependence of $\tau$ on the Poisson ratio $\nu$ is displayed in Fig. (5). We therefore do not plot the transit velocity $\ell/\tau$, which lacks a clear physical interpretation, and instead focus on the transit time $\tau$. A key observation is that $\tau$ varies significantly with $\nu$, and therefore with the effective barrier length, demonstrating that $\tau$ is not independent of $\ell$. This behavior stands in clear contrast with the barrier-length–independent transit times reported in the literature [10, 58, 59].

In such tapered configurations, once a critical barrier length is exceeded ($\ell \gtrsim 20$ mm), no transmission through the evanescent region is observed, so the tunneling-like pathway collapses and no meaningful "tunneling time" can be defined [60]. Consequently, no Hartman-type saturation can occur in this configuration [25]. Lastly, $\tau$ is likewise sensitive to the control-parameter triplet ($\nu$, $\theta$, $f_0$), whose combined action governs the effective barrier length and dictates where, in frequency–space, the evanescent barrier materializes.

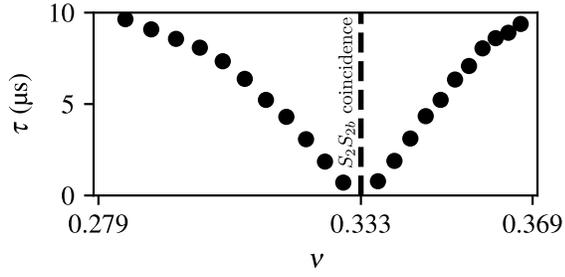

FIG. 5. Evolution of the apparent transit time ($\tau$) within the confined range of Poisson's ratio $\nu$ values around the Dirac cone for which tunneling can occur.

# SUPPLEMENTAL MATERIAL
## Tunneling of Elastic Waves in a Tapered Waveguide

Alexandre Yoshitaka Charau,[1,2] Jérôme Laurent,[1] and Tony Valier-Brasier[2]

[1]*Université Paris-Saclay, CEA, List*, F-91120, Palaiseau, France
[2]*Institut Jean Le Rond ∂'Alembert, Sorbonne Université*,
UMR CNRS7190, 4 Place Jussieu, Paris 75005, France

The Supplemental Material reports: ($i$) The barrier length $\ell$ can be predicted heuristically using a semielliptical function with an asymmetry factor ($a$) that efficiently fits any imaginary wave number loop $k_{S_{2i}}(f)$. The approximate model remains accurate except for the highest $\nu > 0.42$, which lie beyond the range of the effect and exhibit more pronounced asymmetries requiring an additional elliptical deformation factor; nevertheless, the goodness-of-fit (GoF) remains acceptable. The resulting dimensionless relationship $k_{2i}$ depends on the elastic properties of the material, particularly the Poisson's ratio $\nu$. By incorporating the dihedral angle $\theta$ and a chosen frequency $f_0$, this model enables reliable prediction of $\ell$ for any given waveguide cross-section slope. ($ii$) Furthermore, the energy transport behavior for three different Poisson's ratios $\nu$, showing that numerical simulations agree well with experimental results. This energy transport persists for slightly longer barrier lengths at lower $\nu$, but vanishes beyond a critical barrier length, or length of $S_{2i}$, due to increased attenuation from the larger imaginary part of the wave number. ($iii$) Influence of viscosity on the dispersion curves and, consequently, on the effective barrier length $\ell$, thereby clarifying how viscoelastic losses may alter the tunelinglike behavior. ($iv$) We also examine the effect of a reversed gradient within the barrier, for which the $S_2$ mode reappears beyond the evanescent region with a positive wave number rather than connecting to the backward $S_{2b}$ branch, which exhibits a negative phase velocity.

DOI: 10.1103/18p2-8qtp/SM_tunneling.pdf

## SM1. Asymmetric semielliptical expression designed to mimics the imaginary loop

To model the evanescent behavior of the $S_2$ mode, we approximate its imaginary wave number profile, $k_{S_{2i}}(f)$, with a semielliptical function. The loop formed by $k_{S_{2i}}$ as a function of frequency has the shape of an asymmetric ellipse and is strictly bounded between the two cutoff frequencies: the lower cutoff, $f_{c_{S_2}}$, of the $S_2$ mode, and the upper cutoff, $f_{c_{S_{2b}}}$, associated with the backward-propagating $S_{2b}$ branch. Associated dispersion curves [1] are illustrated Figure (S1):

As originally demonstrated by Mindlin, and confirmed through our in-house semi-analytical python code, we intentionally omit the two purely imaginary branches emerging from the midpoint of both imaginary loops ($S_{2i}$), as they signal a polarization change, analogous to that occurring at turning points (cutoff frequencies). The cutoff frequencies are determined by the material's properties and the local thickness of the waveguide ($h$), and are given by:

$$f_{c_{S_2}} = \frac{V_T}{h} \quad \text{and} \quad f_{c_{S_{2b}}} = \frac{V_L}{2h}, \tag{S1}$$

where $V_T$ and $V_L$ denote the transverse and longitudinal wave velocities, respectively. To construct the semielliptical approximation, it is essential to determine the maximum radius of the pure imaginary loop, either the equatorial point of the semi-major axis $k_{S_{2i}}^{\max}$. As depicted in Fig. (S1), the evolution of $k_{S_{2i}}^{\max}$ with Poisson's ratio $\nu$ reveals two distinct regimes, separated at coincidence point $\nu_c = 1/3$. We therefore fit this trend using a polynomial expressions as $k_{S_{2i}}^{max}(\nu) = \sum_{k=0}^{n \leqslant 3} p_k \nu^k$, each tailored to one of the regimes below or above the coincidence point (Dirac cone; *i.e.*, when $f_{c_{S_2}} = f_{c_{S_{2b}}}$):

$$k_{2i}^{\max}(\nu) = \frac{2\pi}{h} \times$$

$$\begin{cases} -4.83\nu^3 + 0.75\nu^2 - 0.45\nu + 0.25 & \text{for } \nu \leqslant 1/3, \\ 6.86(\nu - \nu_c)^2 + 1.81(\nu - \nu_c) & \text{for } \nu \geqslant 1/3. \end{cases} \tag{S2}$$

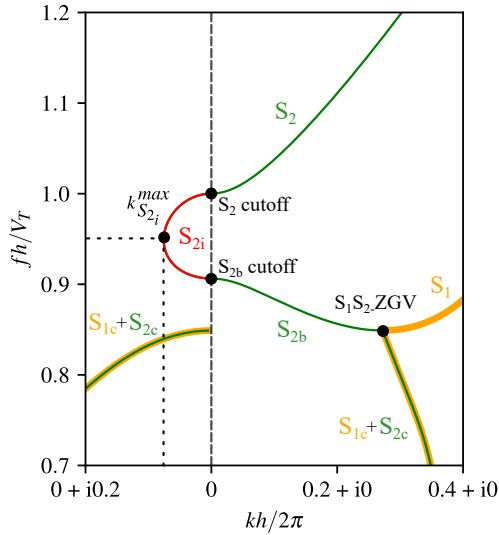

FIG. S1. Dimensionless complex dispersion curves for a metallic plate with Poisson's ratio $\nu = 0.3$.

The result is shown in Fig. (S2), with a GoF close to unity over the ordinarily material range of Poisson's ratio. The green dashed box around the coincidence point (Dirac cone [2, 3]) highlights the region where tunneling-like behavior is likely to occur.

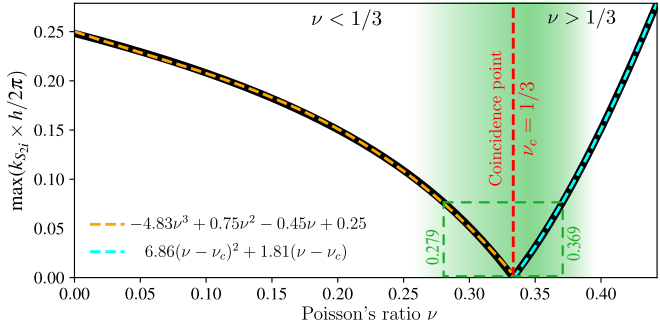

FIG. S2. Evolution of the dimensionless maximum imaginary wave number $k_{S_{2i}}^{\max}$ as a function of Poisson's ratio $\nu$, with $h_i = 1$ mm. The area indicated by the white dashed square corresponds to the tunneling zone, extend $\nu = 0.279$ from $\nu = 0.369$. Polynomial functions of orders 2 and 3 were used to fit the curves in both regimes, respectively, yielding $R^2 \approx 1$.

We then sought an asymmetric semielliptical expression that accurately mimics the imaginary loops of the $S_2$ mode. After a series of mathematical manipulations and rearrangements—starting from the standard equations of an ellipse, as commonly founded in textbooks—we obtained an expression that depends on $\nu$. This was subsequently refined and formalized as a "heuristic model", incorporating the two previously identified dependencies on $\nu$ through the second- or third-degree polynomial fits shown in Fig. (S2). Thus, such a polynomial formulation provides a simple yet accurate representation of the underlying dependence and is inserted into Equ. (S3), resulting in a dimensionless semielliptical expression. This model reproduces well $S_{2i}$ mode when compared with those computed using our in-house Python code, which solves the transcendental (Rayleigh–Lamb) equation in the $(f - k)$ complex plane. Once the maximum value is determined, it is incorporated into the dimensionless semielliptical heuristic expression (Equ. S3), yielding the full frequency dependence of $k_{S_{2i}}(f)$. This expression remains valid across both regimes in the vicinity of the point of coincidence (Dirac cone):

$$k_{S_{2i}} h = k_{S_{2i}}^{\max} \cdot \sqrt{1 - \mathcal{A}^2} \cdot (1 - 2a\mathcal{A}), \quad (S3)$$

where $\mathcal{A} = ((2f/V_T) - (1 + \chi))/(1 - \chi)$ with $\chi = (1/2)\sqrt{2(1-\nu)/(1-2\nu)}$. In Equ. (S3), the asymmetry parameter must be positive or negative below or above the coincidence point, respectively, and have a value close to $a \approx \pm 0.1$. The LM-method applied in Fig. (S3) is used to fit the asymmetry parameter $a$ in the semielliptical expression (Equ. S3). Alternatively, for ease of use, one may fix $a \approx \pm 0.1$, which allows a direct (non-fitted) computation while still maintaining an acceptable goodness-of-fit [4].

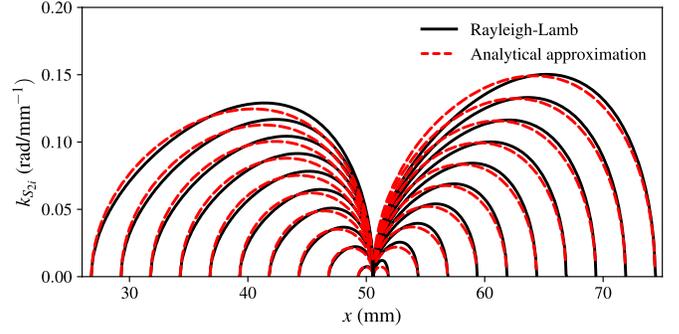

FIG. S3. Comparison of the local imaginary wave number $k_{S_{2i}}$ derived from the Rayleigh–Lamb equation [5] and from our semielliptical heuristic function that mimics the imaginary loops of the $S_{2i}$ mode (Equ. S3). The Levenberg–Marquardt (LM) fitting method is employed. The asymmetry parameter $a$ is treated as a free variable to achieve optimal fit quality. The GoF, evaluated using the $R^2$ statistic, ranges from 0.86 to 0.99.

As reported in Figure (S3), this deformed elliptical relation allows reconstruction of the full shape of the imaginary loop ($S_{2i}$) from $\nu$, including the fitted maximum of $k_{S_{2i}}^{\max}$. It significantly simplifies the integration of $k_{2i}(f)$ used to estimate the barrier length $\ell$, as discussed in the previous section. Besides Poisson's ratio, the barrier length $\ell$ depends on two additional parameters: the selected excitation frequency $f_0$ and the waveguide slope or dihedral angle $\theta = \arctan[(h_1-h_2)/L] \times (180/\pi)$. While $\nu$ governs the onset and extent of evanescent behavior through its effect on the imaginary part of the wave number, the frequency controls the spectral proximity to the mode cutoff, thus determining the tunneling window. Concurrently, the thickness gradient defines the spatial region where the $S_2$ mode enters and exits its evanescent regime, effectively setting the physical width of the tunneling barrier. Steeper gradients produce shorter barriers, whereas shallower gradients extend the tunneling region.

A direct approach to estimate the barrier width $\ell$ from experimental data consists in first measuring the plate thicknesses, for example using a laser rangefinder as employed in our study, and the ZGV resonances by laser-ultrasound technique, as performed in our experiments. This allows determination of the longitudinal and shear wave velocities, as well as the local and absolute Poisson's ratio. Once these elastic parameters $(\nu, V_L, V_T)$ are obtained from the ZGV resonances, the barrier length can be deduced using: $\ell = x(h = V_L/(2f)) - x(h = V_T/f)$. The barrier width $\ell$ can also be verified using Equ. (S3), as detailed in Section (SM1). Solving the inverse problem more generally requires further careful consideration and will be addressed in future work. At this stage, this first approach remains the most reliable and practical strategy.

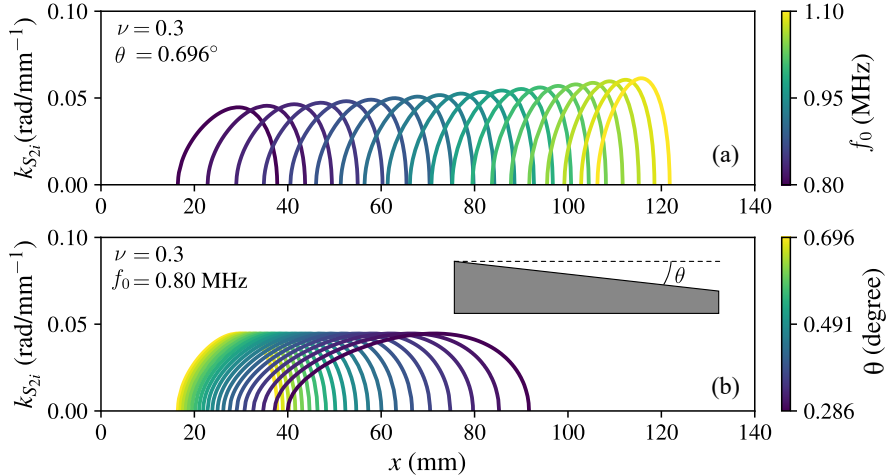

FIG. S4. (a) Evolution of the tunneling barrier for various excitation frequencies ($f_0$) and (b) various soft slope elastic waveguide (dihedral angle $\theta$) achieved with the Equ. (S3).

## SM2. Barrier width $\ell$ tuned by the Poisson's ratio $\nu$

To highlight the different tunneling regimes, we use numerical hybrid SAFE/FE simulations to compute the transmission (**T**) and reflection (**R**) coefficients via the scattering-matrix formalism [6] implemented into the CIVA platform, as also detailed in the Supplementary Material of our previous study [7].

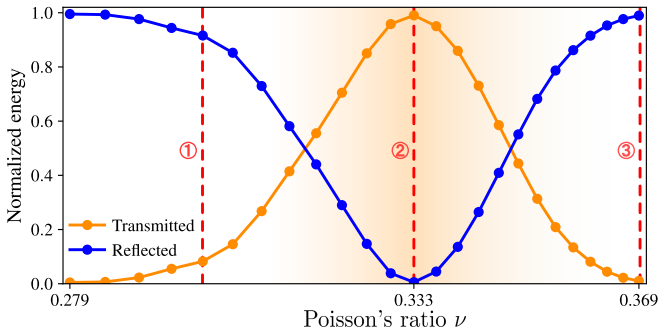

FIG. S5. Normalized energy of the $S_2$ mode: transmitted (**T**) and reflected (**R**) components as functions of Poisson's ratio $\nu$. The red dashed lines indicate three regimes: ① partial reflection/transmission, ② total transmission, and ③ total reflection.

As demonstrated in the Letter, at the working frequency $f_0 = 0.89$ MHz, we compute the normalized transmitted (**T**) and reflected (**R**) energies for Poisson's ratios $\nu$ ranging from 0.279 to 0.369. To capture the distinct **R**/**T** regimes (see Fig. S5) induced by variations in $\nu$, we provide three short animations showing the evolution of the local wave number over time. At $\nu = 0.30$, the wave is partially transmitted and reflected, revealing a regime of moderate tunneling. Strikingly, at $\nu = 1/3$, the barrier becomes effectively invisible, enabling near-perfect transmission via Dirac-cone–mediated propagation. Conversely, increasing $\nu$ to 0.369 transforms the barrier into a near-opaque wall, fully blocking transmission. These animations compellingly demonstrate how tuning $\nu$ acts as a powerful control knob over the effective barrier width and thus over the wave-propagation regime itself.

For ① and ②, we also note that a small portion of the wave packet's energy appears to as-leak through the inhomogeneous branches ($S_{1c}$ and $S_{2c}$) near the saddle point ($f_{S_1 S_2}^{ZGV}$). However, this interpretation is complicated by the dependence of the observed packet length on the sliding window used in signal processing, as well as the significantly increased attenuation along these complex branches. Refer to additional unpublished results discussed in Chapter 4.4.2 of Ph.D work [8]. When the $S_2$ mode reaches the barrier, then the energy is stored-like within the evanescent region for a maximum transit time of $2\tau \gtrsim 20$ $\mu$s (analogously to a capacitor). Beyond the associated critical length $\ell_c \gtrsim 20$ mm (refers to Fig. 4b), the evanescent field joined again the cutoff of $S_2$ mode: the energy reverses direction and the wave recovers its propagative nature along the $S_2$ mode but in the opposite direction like $k' < 0$.

Furthermore, a key observation is that both numerical and experimental results could be further improved by implementing enhanced selective modal excitation. This can be achieved using a Spatial Light Modulator (SLM) [9] or array of transducers to synthesize excitation from a specific segment of the dispersion branch, followed by a 2D inverse Fourier transform to reconstruct the displacement field as a function of space and time for the targeted mode [10].

## SM3. Impact of viscoelastic dissipation on the barrier length $\ell$

Finally, to elucidate how viscoelastic losses may affect the tunnelinglike behavior, we analyze the influence of viscosity on the dispersion curves and, consequently, on the effective barrier length $\ell$.

When viscosity, whose effect typically follows a frequency-dependent power law [5], is incorporated, for example through a Kelvin–Voigt model, the dispersion

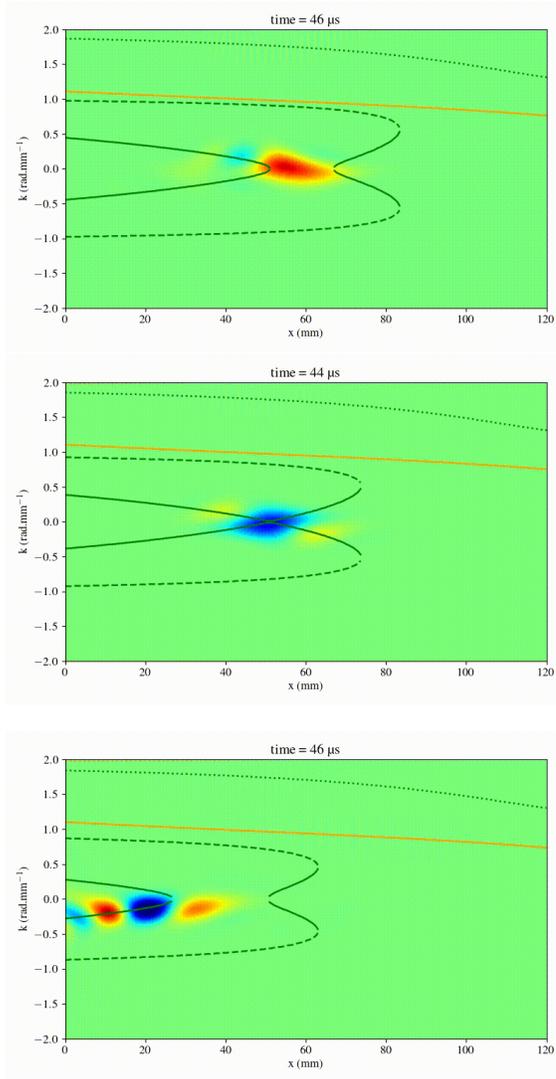

① **Tunneling regime** with $\nu = 0.300$: the barrier corresponds to a region where guided modes cannot exist at the given frequency; the wave becomes evanescent: it decays exponentially without real propagation. If the barrier is sufficiently thin, the wave can recover its propagative nature on the other side. Figure SM6a (static frame). The animated version is available as an ancillary file on arXiv.

② **Classical regime** at Dirac cone with $\nu = 0.333$: quasi-free propagation occurs without significant attenuation, even through the perturbed zone; at the intersections of $f_{c_{S_1}}$ and $f_{c_{S_2}}$. Reflection is minimal and transmission is nearly complete. Figure SM6b (static frame). The animated version is available as an ancillary file on arXiv.

③ **Total reflection** with $\nu = 0.369$: the evanescent wave nearly tunnels through but is ultimately blocked at the barrier's far end. For $\nu > 1/3$, the $S_1$ and $S_2$ modes swap labels under physical mode numbering, based on the number of displacement nodes across the plate thickness. Figure SM6c (static frame). The animated version is available as an ancillary file on arXiv.

FIG. S6. Movies of numerical spatial spectrograms at different regimes of tunneling and interaction with the evanescent region for three different Poisson's ratios. From top to bottom, solid lines indicate modes $A_0$, $S_0$, $\pm S_1$, $\pm S_2$, and $S_{2b}$. These animations offer a striking time-resolved visualization of the modal energy redistribution that enables tunneling through the evanescent barrier. In such configuration, they thus offer an unambiguous and physically transparent demonstration of the respective roles played by the participating mode.

characteristics are primarily altered near the turning and saddle points associated with ZGV modes, where the wavenumber becomes complex rather than purely imaginary. In these regions, if the attenuation is very high, well beyond the values measured in metallic plates such as aluminum, both the cutoff frequencies and the extent of the purely imaginary $S_{2i}$ loop are modified, and the modal polarization ceases to remain orthogonal, varying progressively in the vicinity of these points.

In our case, the viscoelastic attenuation has been experimentally measured to be approximately 1 dB/m at 5 MHz. For such low losses, the effect on both the turning points and the ZGV modes is negligible, as illustrated in Fig. (S7a). It may therefore be inferred that increasing attenuation shortens the imaginary loop and consequently reduces the effective barrier width $\ell$ [Fig. S7(b)]. For soft or polymer-based waveguides, however, this assumption no longer holds, and viscoelastic damping must be explicitly accounted for, as done by Delory *et al.* [11].

In other words, when the evanescent barrier supports a complex wave number with a nonzero real part, *i.e.*, an inhomogeneous mode exhibiting simultaneous exponential decay and spatial oscillation, the regime differs fundamentally from the purely evanescent tunneling scenario, where $k$ is strictly imaginary. Nevertheless, it represents a physically meaningful and highly intriguing case, opening a particularly stimulating direction for future investigations. More elaborate geometrical designs—such as modified gradients, multi-segment or piecewise-linear thickness profiles, or locally non-monotonic variations—may naturally give rise to complex-valued wave number solutions within the bar-rier. Studying these configurations would provide access to a broader class of inhomogeneous states and offer insight into how they shape energy transport, phase evolution, and mode interactions in elastic waveguides. Such regimes may reveal phenomena that extend beyond the strictly evanescent tunneling process reported here.

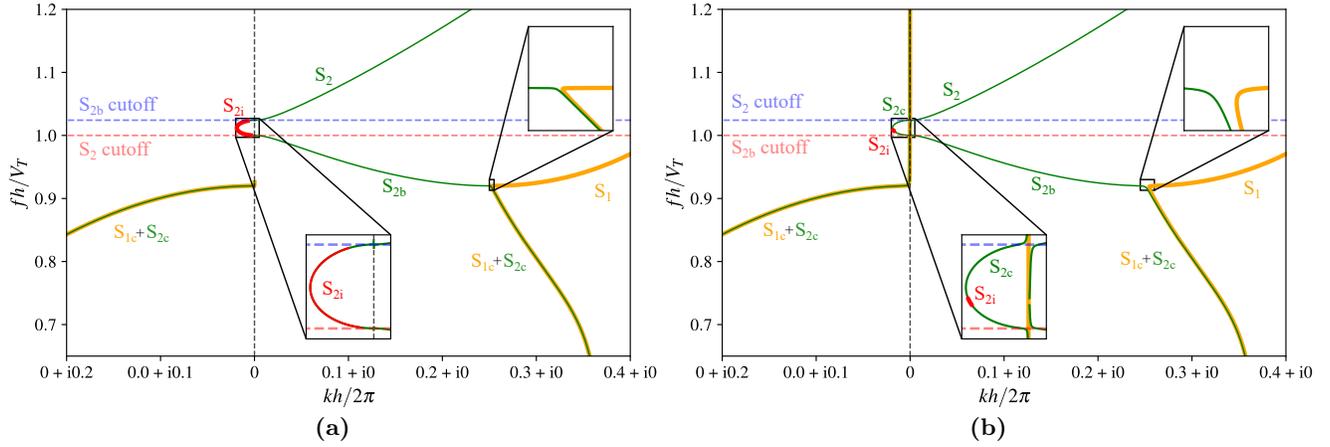

FIG. S7. Dimensionless complex dispersion curves were calculated by incorporating a linearly frequency-dependent viscoelastic attenuation, calibrated to approximately (a) 1 dB/m or (b) 50 dB/m at 5 MHz. (a) For such low attenuation, the turning points remain unaffected, and the ZGV branches preserve their characteristic stick-like behavior. (b) For higher attenuation levels, unrealistic for metallic plates, the curvature at the turning points increases, and the $S_1$ and $S_2$ branches progressively open up or move away from each other.

### SM4. Reversed gradient within the barrier

In the configuration previously considered, the waveguide is tapered such that the $S_2$ mode reaches its cutoff thickness and becomes evanescent, before re-emerging as the $S_{2b}$ mode with a negative wave number. This backward branch carries energy forward, *i.e.*, with positive group velocity $V_g^{S_{2b}}$, while exhibiting a negative phase velocity $V_\varphi^{S_{2b}}$, yielding a distinctive tunneling scenario.

If however, the thickness gradient is reversed within the barrier, the $S_2$ mode would still become evanescent as it approaches its cutoff thickness, but would re-emerge beyond the barrier not as $S_{2b}$, but as the original $S_2$ mode with a positive wave number. In this case, an evanescent region still forms, yet the transmitted wave recovers both positive phase and group velocities. This scenario is illustrated in Fig. (S8), which shows snapshots of the spatial spectrogram at successive times: the incident $S_2$ mode becomes evanescent within the barrier and then reappears with a positive wave number, fully consistent with this description. This leads to a distinct tunnelinglike mechanism: the mode on both sides of the barrier remains $S_2$, and the transition occurs across a symmetric thickness profile. Energy transfer still proceeds through an evanescent region, but the symmetry ensures that the transmitted mode is forward-propagating, unlike in our tapered configuration, where the transmitted state necessarily connects to the backward $S_{2b}$ branch. This reversed-gradient configuration is conceptually close to that analyzed by Nielsen and Peake [12] in the context of acoustic-wave tunneling in slowly varying axisymmetric flow ducts. However, this last point highlights a key distinction between our results and earlier studies on tunnelinglike phenomena.

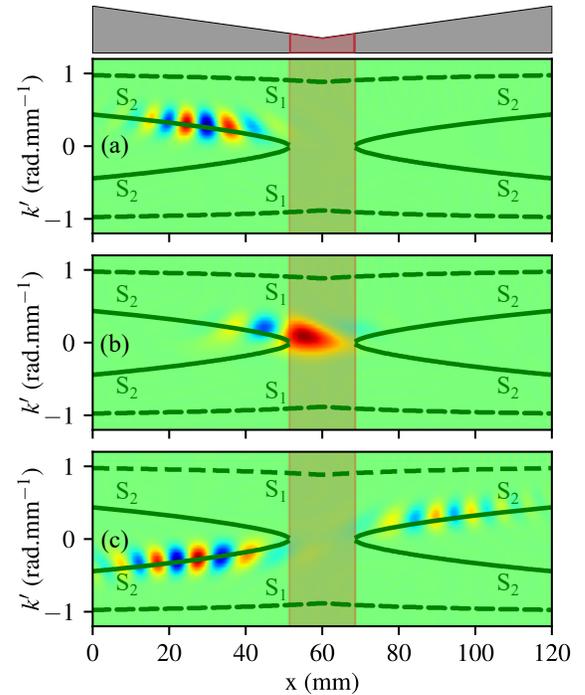

FIG. S8. Numerical spatial spectrogram $S_{f_0}(x,t,k)$ of the $S_2$ mode at different stages for $\nu = 0.3$: (a) before tunnelinglike behavior at 37 $\mu$s; (b) interaction with the evanescent region at 47 $\mu$s; (c) after tunnelinglike behavior at 63 $\mu$s.

---

[1] R. D. Mindlin, *An introduction to the mathematical theory of vibrations of elastic plates*. World Scientific Publishing Co. Pte. Ltd. (2006).
[2] A. Maznev, *J. Acoust. Soc. Am.* **135**, pp. 577–580 (2014).